\documentclass{aa}
\usepackage{graphics} 
\begin{document}
\title{ISOCAM Mid-InfraRed Detection of HR 10: A Distant Clone of Arp
220 at $z$=1.44}

\author{\bf D.~Elbaz\inst{1,}\inst{2,}\inst{3}
\and H.~Flores\inst{1} 
\and P.~Chanial\inst{1}
\and I.F.~Mirabel\inst{1,4} 
\and D.~Sanders\inst{5}
\and P.-A.~Duc\inst{1}
\and C.J.~Cesarsky\inst{6}
\and H.~Aussel\inst{5}}
\offprints{D. Elbaz}
\institute{
CEA Saclay/DSM/DAPNIA/Service d'Astrophysique, Orme des
Merisiers, 91191 Gif-sur-Yvette C\'edex, France
\and 	Department of Physics, University of California, Santa Cruz, CA 95064, USA
\and 	Department of Astronomy \& Astrophysics, University of California, Santa Cruz, CA 95064, USA
\and    Instituto de Astronom\'\i a y F\'\i sica del Espacio, CONICET, 1428 Ciudad Universitaria, Buenos Aires, Argentina.
\and	Institute for Astronomy, University of Hawaii, 2680 Woodlawn Drive, 96822 Honolulu, USA
\and    European Southern Observatory, Karl-Schwarzchild-Strasse, 
	2 D-85748 Garching bei M\"unchen, Germany \\
\email{delbaz@cea.fr}
}

\date{received ; accepted }
%
\abstract{We report the detection of the extremely red object (ERO),
HR 10 ($I-K\sim 6.5$, $z$=1.44), at 4.9 and 6.1\,$\mu$m (rest-frame)
with ISOCAM, the mid-infrared (MIR) camera onboard the Infrared Space
Observatory (ISO). HR 10 is the first ERO spectroscopically identified
to be associated with an ultra-luminous IR galaxy (ULIG) detected in
the radio, MIR and sub-millimeter. The rest-frame spectral energy
distribution (SED) of HR 10 is amazingly similar to the one of Arp
220, scaled by a factor $3.8\pm1.3$. The corresponding 8-1000\,$\mu$m
luminosity ($\sim$ 7$\times$10$^{12}~h_{70}^{-2}$ $L_{\sun}$) translates
into a star formation rate of about $1200~h_{70}^{-2}$ $M_{\odot}$
yr$^{-1}$ if HR 10 is mostly powered by star formation. We address the
key issue of the origin of the powerful luminosity of HR 10,
i.e. starburst versus active galactic nucleus (AGN), by using the
similarity with its closeby clone, Arp 220.   } \maketitle

\section{Introduction} 
\label{introduction}
\begin{figure*}[t!]
\resizebox{\hsize}{!}{\includegraphics{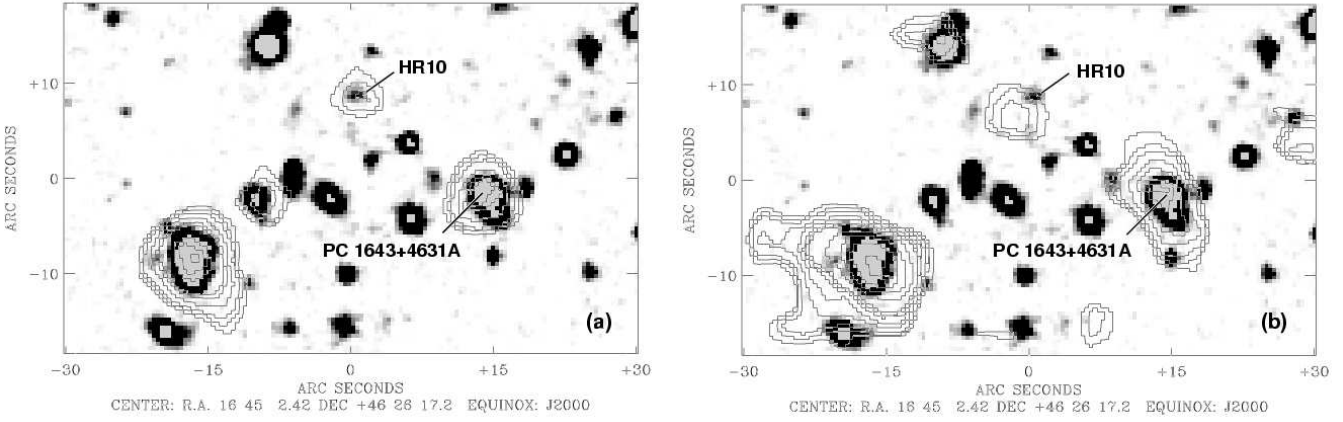}}
\caption{MIR contours over $I$-band image (WHT archives); N: up, E:
left. {\bf a)} 15\,$\mu$m contours at 2.5 to 5-$\sigma$, with a
0.5-$\sigma$ step. {\bf b)} 12\,$\mu$m contours at $S/N$=1.6, 1.9,
2.2, 2.8, 3.5, 4.1, 6, 9, 11.}
\label{FIG:cam}
\end{figure*}
HR 10 (or ERO J164502$+$4626.4, Dey et al. 1999), is the first and
presently only Extremely Red Object (ERO, usually defined as galaxies
with $I-K > 4$) known to be associated with the class of
ultra-luminous infrared galaxies (ULIGs, $L_{\rm IR}$=
$L$(8-1000\,$\mu$m)$\geq$ 10$^{12}~L_{\sun}$).  It was detected by Hu
\& Ridgway (1994, hereafter HR) together with another ERO (HR 14 or
ERO J164457$+$4626.0) in the field of the QSO PC 1643+4631A ($z$=
3.79). HR initially suggested that both galaxies with extreme colors
($I-K > 6$) could be distant ellipticals lying at $z\sim 2-3$. More
generally, deep near IR (NIR) surveys indicate that EROs present the
same clustering properties (Daddi et al. 2000, McCarthy et al. 2001)
and surface brightness distribution (Moriondo, Cimatti \& Daddi 2000)
as elliptical galaxies. But dusty starbursts being potential
progenitors of local ellipticals, they may also show similar
clustering properties and some local examples, as NGC 7252 (Hibbard et
al. 1994) or Arp 220 (Scoville et al. 2000), already show a de
Vaucouleurs luminosity profile typical of ellipticals. High resolution
NIR imagery and spectroscopy with the Keck telescopes (Graham \& Dey
1996) revealed that HR 10 was a moderately distant ($z=1.44$) galaxy
with an asymmetric morphology and H${\alpha}$ in emission. Another
evidence against HR 10 being an early-type galaxy is its strong
sub-millimeter luminosity as measured with SCUBA at the JCMT (Cimatti
et al. 1998, Dey et al. 1999). The detection in HR 10 of a large CO
luminosity, hence molecular hydrogen mass, was recently reported by
Andreani et al. (2000) and presented as evidence favoring a star
formation origin for the bulk of the IR luminosity rather than an
AGN. However, a large gas mass may not only feed star formation but
also gas accretion onto an AGN (see Papadopoulos et al. 2001). Another
test for the presence of an AGN in a dusty galaxy is the warm over
cold dust, i.e mid IR (MIR, 3-40\,$\mu$m) over far IR (FIR,
40-300\,$\mu$m), luminosity ratio as well as the shape of the MIR
spectrum. We present the first detection of HR 10 in two MIR bands of
ISOCAM onboard ISO corresponding to the rest-frame 3.3-6.1 and
4.9-7.4\,$\mu$m wavelength ranges. A comparison to local galaxies SEDs
is presented. We show that the SED of HR 10 is a scaled version of
that of the closest ULIG, Arp 220. We discuss the origin of the
luminosity of HR 10 on the basis of this similarity.

Throughout this paper, we will assume $H_o$= 70 km
s$^{-1}$ Mpc$^{-1}$, $\Omega_{\rm matter}=~0.3$ and
$\Omega_{\Lambda}=~0.7$. For this cosmology, the luminosity distance
of HR 10 ($z=1.44$) is 10373 Mpc.
\section{Observations}
\label{Observations} 
Deep images of a field of $13'\times13'$ centered on the QSO PC
1643$+$4631A ($z=3.79$) were obtained with the ISOCAM broadband
filters LW3 (12-18\,$\mu$m, centered at 15\,$\mu$m) and LW10
(8-15\,$\mu$m, centered at 12\,$\mu$m) for a total integration time of
16 and 13 minutes respectively. The 15\,$\mu$m image results from the
coaddition of two overlapping mosaics slightly rotated one with
respect to the other resulting in an improved spatial resolution
(2$\arcsec$ pixels). The 12\,$\mu$m image being made of one single
mosaic of 6$\arcsec$ pixels suffers from a lower spatial resolution. A
given position of the sky was observed by 18 (20) different pixels at
15\,$\mu$m (12\,$\mu$m), resulting in a better flat-fielding and
correction of cosmic ray impacts. The data reduction and source
extraction was done with PRETI (Pattern REcognition Technique for
ISOCAM data, Starck et al. 1999). The gaussian noise (photon plus
detector noise) in the 15\,$\mu$m (12\,$\mu$m) image is 58\,$\mu$Jy
(31\,$\mu$Jy). This field was also observed with ISOPHOT onboard ISO
at 90 and 170\,$\mu$m (P.I. R.Ivison). The total exposure time per sky
position is 11.5 and 6.6 minutes respectively. HR 10 is not detected
with a 5-$\sigma$ upper limit of $S_{\nu}\sim 200$ mJy in both bands.

On Fig.~\ref{FIG:cam}, the MIR contours are overlayed on the $I$-band
image of a field containing both HR 10 and PC 1643+4631A. Both
galaxies are detected in each MIR band with the flux densities given
in Table~\ref{TAB:Snu}. The uncertainty on the flux density related to
the correction for the transient behavior of the detectors and to the
pixel size relative to the point spread function is given in a second
line in italics. While the 15\,$\mu$m contours are centered on the
optical positions, the 12\,$\mu$m contours of HR 10 present an offset
of 4$\arcsec$. The astrometry was first calculated over the whole
13$\arcmin$ field using several stars from the US Naval Observatory
Catalog which were detected in the MIR. A relative astrometric
correction was then applyied to check the position of the MIR
detection at the optical position of HR 10 by using a set of six
objects included in a field of 2$\arcmin\times$2$\arcmin$ centered on
HR 10 and including PC 1643+4631A. The probability of a chance
association with an optical object with an $I$-band magnitude lower
than $I$ within a distance $d$ was estimated using the following
formula assuming a Poissonian distribution of sources:
$P=1-\exp[-n(I)\pi d^2]$. The offset of 4$\arcsec$ for the 12\,$\mu$m
detection is attributed a probability of a chance association with the
optical position of HR 10 of 20\,$\%$. This is not surprising because
of the lower spatial resolution of the 12\,$\mu$m image (6$\arcsec$
pixels). However the probability of a random association of the
12\,$\mu$m source within 4$\arcsec$ of the 15\,$\mu$m detection of HR
10 is equal to 0.1\,$\%$ ($n$[0.2 mJy at 15\,$\mu$m]= 0.8$\pm$0.1
$\arcmin^{-2}$). 
In order to quantify the risk of detecting ``ghost'' sources produced
by cosmic ray residuals in the MIR images, we performed Monte-Carlo
simulations by inserting fake sources in real datacubes. Down to a
flux density limit of $S_{\nu}\sim 100~\mu$Jy, we find that about 6
and 12\,$\%$ of the sources are false in the 15 and 12\,$\mu$m images
respectively. However, we were able to reduce the fraction of ghost
sources to zero in the 15\,$\mu$m image by requiring that a source be
detected in both mosaics (before coaddition) at a lower detection
level. Only above 0.3 mJy do we reach the zero probability of a false
detection in the 12\,$\mu$m image, but again the probability of a
random association of a spurious 12\,$\mu$m source with the 15\,$\mu$m
detection of HR 10 is negligible. The upper limits on HR 14 ($I-K=$
6.2, see Table~\ref{TAB:Snu}) imply that it cannot be a ULIG unless it
is more distant than $z\sim 1.8$. PC 1643$+$4631A is one of the most
distant sources detected in the MIR ($z=3.79$ corresponds to an age of
the universe of only 1.6 Gyr) and belongs to the class of ULIGs, with
rest-frame luminosities of: $\nu L_{\nu}[3.1 \pm 0.63\,\mu{\rm
m]}=(2.3\pm0.5)\times 10^{12}~h_{70}^{-2}~L_{\sun}$ and $\nu
L_{\nu}[2.5 \pm 0.84\,\mu {\rm m}]=(1.2\pm0.8)\times
10^{12}~h_{70}^{-2}~L_{\sun}$.

\begin{table}
\begin{tabular}{|l|rlcrlc|}
\hline
	& $S_{15\,\mu m}$  & $S/N$  & $P$ & $S_{12\,\mu m}$   & $S/N$ & $P$    \\
	&  ($\mu$Jy)       &      &($\%$)&  ($\mu$Jy)     &     &($\%$)\\
\hline
HR 10    & 203$\pm$58       & 3.5  & 2.5  & 85$\pm$31      & 2.2 & 20   \\
        & $\pm${\it 22}    &      &      & $\pm${\it 39}  &     &      \\
PC1643A & 330$\pm$58       & 5.6  & 0.8  & 140$\pm$31     & 3.6 & 2.0  \\
        & $\pm${\it 12}    &      &      & $\pm${\it 39}  &     &      \\
HR 14	& $< 290$          &  5   &  -   & $< 135$        & 5   & -    \\
\hline
\end{tabular}
\caption{MIR flux densities at 15 and 12\,$\mu$m, signal-to-noise
ratios and probability of chance association.}
\label{TAB:Snu}
\end{table}
\section{Discussion: Nature of HR 10}
\label{Discussion} 
\begin{figure*}
\resizebox{\hsize}{!}{\includegraphics{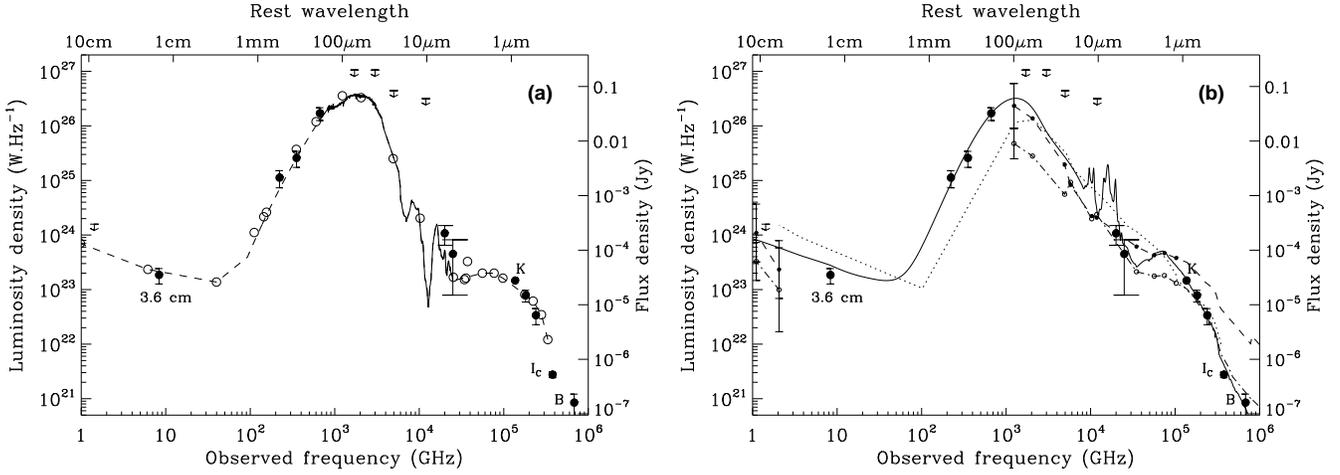}}
\caption{\em Rest-frame SED of HR 10 (filled circles, Dey et al. 1999)
including our MIR data. {\bf a)} comparison to the SED of Arp 220
normalized by a factor 3.8.  Upper limits from IRAS ($\lambda_{\rm obs}$=
25, 60\,$\mu$m) and ISOPHOT (90, 170\,$\mu$m). Plain line: combined
spectra from ISO-LWS (43-185\,$\mu$m, Fischer et al. 1999) and
ISOCAM-CVF (5-18\,$\mu$m, Charmandaris et al. 1999). Dashed line:
interpolation between observed magnitudes: $UV$ (RC3), $BI$ (Surace et
al. 2000), $RcJHKL$ (Kim 1995), IRAS (FSC), ISOPHOT (Klaas et
al. 1997), sub-millimeter (Rigopoulou et al. 1996), radio
(Anantharamaiah et al. 2000). {\bf b)} comparison to the mean SEDs of
a ``high reddening starburst'' (dashed line, filled circles) and a
Seyfert 2 (dash-dotted line, empty circles) from Schmitt et
al. (1997), and to Mrk 231 (dotted line, Ivison et al. 1998).}
\label{FIG:sed_hr10}
\end{figure*}

The least-square fit of the SED of HR 10 using the SED of Arp 220
(Fig.~\ref{FIG:sed_hr10}a) illustrates the amazing similarity of both
spectra. The 8-1000\,$\mu$m luminosity of Arp 220 is $L_{\rm
IR}=~1.8\times10^{12}~h_{70}^{-2}~L_{\sun}$, for a distance of
$d$=~81.8 $h_{70}^{-1}$ Mpc, corrected from the Virgo infall
($cz$=~5439+206 km s$^{-1}$). The resulting normalization factor of
$3.8\pm1.3$ (1-$\sigma$) implies an IR luminosity of $L_{\rm
IR}=~(6.8\pm2.3)\times10^{12}~h_{70}^{-2}~L_{\sun}$ for HR 10. If the
dust was mostly heated by massive young stars, then this IR luminosity
would translate into a star formation rate ($SFR$) of about
$(1170\pm396)~h_{70}^{-2}~M_{\sun}$ yr$^{-1}$, for a 10-100 Myr
continuous burst, solar abundance and a Salpeter IMF
($SFR~[M_{\sun} {\rm yr}^{-1}]=~1.72\times10^{-10}~L_{\rm IR}$, Kennicutt
1998).

A consistent normalization factor also applies for the molecular gas
mass of HR 10, estimated from its CO luminosity ($M_{\rm H_2}$=
1.2$\times$10$^{11}$ $M_{\sun}$, Andreani et al. 2000), which is 2.9
times larger than in Arp 220 ($M_{\rm H_2}$[Arp 220]=
4.1$\times$10$^{10}$ $M_{\sun}$, Scoville, Yun \& Bryant 1997). If the
molecular gas mass over dust mass ratio of HR 10 is the same as the
one measured in the center of Arp 220, i.e. $M_{\rm gas}/M_{\rm
dust}\sim~400$ (Scoville, Yun \& Bryant 1997), then we obtain $M_{\rm
dust}[{\rm HR 10}]\sim~4\times10^{8}~h_{70}^{-2}~M_{\sun}$, in
agreement with the one estimated by Dey et al. (1999,
$\sim~7\times10^{8}~h_{70}^{-2}~M_{\sun}$) using a black body fit to
the FIR part of the SED with $T_{\rm dust}\sim$ 40 K.

The SED of Arp 220 is limited to wavelengths above the U-band
(3550\,$\AA$) which corresponds to 8670\,$\AA$ in the observed frame
of HR 10. In Fig.~\ref{FIG:sed_hr10}b, we have fitted the SED of HR 10
with the model STARDUST2 (Chanial et al., in preparation) from the
observed B-band (1800\,$\AA$ in the rest-frame) to the radio. If we
assume that there is only one component responsible for both the
optical and IR light in HR 10, then the best-fit is obtained for a
visual extinction of $A_{\rm V}$ = 3 mag ($\pm$ 0.2 mag, extinction law
$A(\lambda)/A_{\rm V}$ from Calzetti et al. 2000), with an age of 0.2 Gyr
at the time of the observation and a characteristic timescale for the
starburst of $\tau$= 95 Myr ($SFR$= $M_{\rm gas}(t)/\tau$). The optical-UV
part of the spectrum alone can be fitted with an $A_{\rm V}$= 2-2.8, an
age of 2-4 Gyr and $\tau$= 0.4-0.9 Gyr, resulting in $L_{\rm
IR}\sim~8\times10^{10}~h_{70}^{-2}~L_{\sun}$ which does not fit the
MIR and sub-millimeter part of the SED. However, detailed studies of
local luminous IR galaxies have shown that the region from which the
bulk of the IR luminosity arises can contribute very weakly to the
optical light, e.g. the Antennae galaxies (Mirabel et al. 1998). In
such a scenario, a second component with a larger $A_{\rm V}$ would
contribute dominantly to the IR regime and nearly not to the optical
part. In both cases, we obtain an IR luminosity and $SFR$ within the
error bars associated with the fit using Arp 220. The main difference
between the model and Arp 220's SED is that the FIR over MIR
luminosity ratio of Arp 220 is higher. The fit of Arp 220's SED
requires an $A_{\rm V}>$ 30 mag for the component responsible for the bulk
of the IR light.

The key issue remains to determine whether the bulk of the luminosity
of HR 10 is powered by star formation or accretion by a black hole.
Two other EROs have been spectroscopically identified and detected in
the MIR and radio (ISO J1324-2016, Pierre et al. 2001, and ERO
J164023+4644, Smith et al. 2001). The upper limits established for
their bolometric IR luminosities in the absence of a detection in the
FIR or sub-millimeter are consistent with their belonging to the class
of ULIGs too. NIR spectroscopy in both galaxies favor the
presence of an AGN. In the case of HR 10, the width of the observed
H$_{\alpha} +[$NII$]$ emission feature or the [NII]/H${\alpha}$ ratio
are too uncertain to be used as an indication of the presence of an
AGN (Graham \& Dey 1996). The H${\alpha}$ line is strongly affected by
dust extinction ($SFR$[H${\alpha}$]$\sim$~80 $M_{\sun}$ yr$^{-1}$
only).

Most authors favor the starburst hypothesis as a dominant source of
energy in Arp 220: its MIR spectrum up to 40\,$\mu$m shows no evidence
for high ionization lines expected for AGNs (Sturm et al. 1996), its
``IR excess'' ($L_{\rm IR}/L[{\rm Ly\alpha}]\sim24$, Anantharamaiah et
al. 2000) is much lower than for AGNs ($\sim 45-65$) and typical of
starburst galaxies ($\sim 12-45$, Genzel et al. 1998), its radio
emission is produced by several compact sources (Smith et al. 1998),
the ratio of aromatic features over MIR continuum (Genzel et al. 1998)
and the slope of the MIR continuum (Laurent et al. 2000) are typical
of starbursts. More recently, Haas et al. (2001) suggested that the
MIR luminosity of Arp 220 could be underestimated because of dust
extinction in the MIR and that after dereddening, its FIR over MIR
luminosity ratio would be closer to the one for AGNs. But the
dereddening factor varies by a factor five depending on the dust
geometry assumed. Finally, the flat 2-10 keV hard X-ray spectrum of
Arp 220 implies that in order to be mostly powered by an AGN, it would
need to be Compton thick with a column density larger than 10$^{25}$
cm$^{-2}$ (Iwasawa et al. 2001).

In Fig.~\ref{FIG:sed_hr10}b, the SED of HR 10 is compared to the mean
SEDs of a Seyfert 2 (Sy2) and of a starburst with high reddening
(SBH), from Schmitt et al. (1997), normalized to the 15\,$\mu$m
luminosity density of HR 10. The 1-$\sigma$ error bars corresponding
to the distribution of the 15 SBH and 15 Sy2 galaxies are shown in the
radio and at $\lambda_{\rm rest}$= 100\,$\mu$m. Sub-millimeter data
are missing in these SEDs but they should decrease in luminosity
density above 100\,$\mu$m as in local galaxies, e.g. the Seyfert 1 Mrk
231 ($d$=~180.9 $h_{70}^{-1}$ Mpc, $L_{\rm
IR}=~3.5\times10^{12}~h_{70}^{-2}~L_{\sun}$, dotted line). Without MIR
data, Mrk 231 was a candidate template for HR 10 (see Dey et
al. 1999). Although not as red as HR 10, the SBH fits the MIR-FIR
region of HR 10 while the FIR luminosity of the Sy2 is about ten times
fainter.

Finally, even if the presence of a combination of an AGN and a starburst
is still an option for both HR 10 and Arp 220, most studies favor a
dominant contribution from star formation to their IR luminosities,
implying that HR 10 presents the largest $SFR$ known at present. This
study clearly shows the need for direct MIR and FIR/sub-millimiter
observations of distant dusty galaxies to improve our understanding of
a population of objects which plays a major role in galaxy evolution
(Chary \& Elbaz 2001), hence emphasizes the importance of the next
generation IR satellites to come, i.e. SIRTF, FIRST and NGST.

\begin{acknowledgements}
We wish to thank Guilaine Lagache for the reduction of the ISOPHOT
images.  DE wishes to thank the American Astronomical Society for
their support through the Chretien International Research Grant, and
Joel Primack \& David Koo for helping funding this research with the
NASA grants NAG5-8218 and NAG5-3507.
\end{acknowledgements}

\end{document}